\documentclass[10pt,conference]{IEEEtran}

\usepackage{algorithmic}
\usepackage{tikz}
\usepackage{amsmath}
\usepackage{pgfplots}
\ifCLASSINFOpdf
\else
\fi
\usepackage{amsmath}
\begin{document}
%
\title{\textbf{\large Efficient Twitter Sentiment Classification using Subjective Distant Supervision}}

\author{\IEEEauthorblockN{Tapan Sahni\IEEEauthorrefmark{1}, Chinmay Chandak\IEEEauthorrefmark{1}, Naveen Reddy \IEEEauthorrefmark{1}, Manish Singh\IEEEauthorrefmark{2}}
\IEEEauthorblockA{Indian Institute of Technology, Hyderabad}
Email: \{cs13b1030, cs13b1011, cs13b1010, msingh\}@iith.ac.in}


%


\maketitle

\begin{abstract}
As microblogging services like Twitter are becoming more and more influential in today's globalized world, its facets like sentiment analysis are being extensively studied.  We are no longer constrained by our own opinion.  Others' opinions and sentiments play a huge role in shaping our perspective. In this paper, we build on previous works on Twitter sentiment analysis using Distant Supervision. The existing approach requires huge computation resource for analyzing large number of tweets. In this paper, we propose techniques to speed up the computation process for sentiment analysis. We use tweet subjectivity to select the right training samples. We also introduce the concept of EFWS (Effective Word Score) of a tweet that is derived from polarity scores of frequently used words, which is an additional heuristic that can be used to speed up the sentiment classification with standard machine learning algorithms. We performed our experiments using 1.6 million tweets. Experimental evaluations show that our proposed technique is more efficient and has higher accuracy compared to previously proposed methods. We achieve overall accuracies of around 80\% (EFWS heuristic gives an accuracy around 85\%) on a training dataset of 100K tweets, which is half the size of the dataset used for the baseline model. The accuracy of our proposed model is 2-3\% higher than the baseline model, and the model effectively trains at twice the speed of the baseline model. 
\end{abstract}

%
\IEEEpeerreviewmaketitle

\section{Introduction}
A lot of work has been done in the field of Twitter sentiment analysis till date. Sentiment analysis has been handled as a Natural Language Processing task at many levels of granularity. Most of these techniques use Machine Learning algorithms with features such as unigrams, n-grams, Part-Of-Speech (POS) tags. However, the training datasets are often very large, and hence with such a large number of features, this process requires a lot of computation power and time. The following question arises: What to do if we do not have resources that provide such a great amount of computation power? The existing solution to this problem is to use a smaller sample of the dataset. For sentiment analysis, if we train the model using a smaller randomly chosen sample, then we get low accuracy [16, 17]. In this paper, we propose a novel technique to sample tweets for building a sentiment classification model so that we get higher accuracy than the state-of-the-art baseline method, namely Distant Supervision, using a smaller set of tweets. Our model has lower computation time and higher accuracy compared to baseline model. 

Users often express sentiment using subjective expression. Although objective expressions can also have sentiment, it is much rare. Determining subjectivity is quite efficient compared to determining sentiment. Subjectivity can be determined for individual tweets. But to do sentiment classification, we need to build a classification model with positive and negative sentiment tweets. The time to train a sentiment classification model increases with the increase in the number of training tweets. In this paper, we use tweet subjectivity to select the best training tweets. This not only lowers the computation time but also increases the accuracy because we have training data with less noise. Even the created features will be more relevant to the classification task. The computation cost will reduce due to small training data size and better set of features. Thus if users do not have enough computational resources, they can filter the training dataset using a high value of subjectivf
ity threshold. This ensures reliable prediction on a smaller training dataset, and eventually requires less computational time. The above approach, and some of the intricacies that invariably seep in, need to be considered, and are described in the later sections of the paper. In this paper we also integrate a lot of meticulous preprocessing steps. This makes our model more robust, and hence leads to higher accuracy.

Along with the machine learning algorithms being used, we use a heuristic-based classification of tweets. This is based on the EFWS of a tweet, which is described in later sections. This heuristic basically takes into account the polarity scores of frequently used words in tweets, and is able to achieve around 85\% accuracy on our dataset, hence boosting the overall accuracy by a considerable amount.

Our training data consists of generic (not topic-specific) Twitter messages with emoticons, which are used as noisy labels. We show that the accuracy obtained on a training dataset comprising 100K tweets, and a test dataset of 5000 tweets gives an accuracy of around 80\% on the following classifiers: Naive Bayes, RBF-kernel Support Vector Machine, and Logistic Regression. Our model takes roughly half the time to train and achieves higher accuracy (than the baseline model) on all the classifiers. Because the amount of training time is expected to increase exponentially as the training data increases, we expect our model to outperform (in terms of higher accuracy) the baseline model at a speed which is at least twofold the speed of the baseline model on larger datasets.  

\section{Related Work}
There has been a large amount of prior research in sentiment analysis of tweets. Read [10] shows that using emoticons as labels for positive and sentiment is effective for reducing dependencies in machine learning techniques. Alec Go [1] used Naive Bayes, SVM, and MaxEnt classifiers to train their model. This, as mentioned earlier, is our baseline model. Our model builds on this and achieves higher accuracy on a much smaller training dataset.\\
Ayushi Dalmia [6] proposed a model with a more involved preprocessing stage, and used features like scores from Bing Liu’s Opinion Lexicon, and number of positive, negative POS tags. This model achieved considerably high accuracies considering the fact that their features were the not the conventional bag-of-words, or any n-grams. The thought of using the polarity scores of frequently used tweet words (as described in our EFWS heuristic) was inspired from this work. [14] created prior probabilities using the datasets for the average sentiment of tweets in different spatial, temporal and authorial contexts. They then used a Bayesian approach to combine these priors with standard bigram language models. \\ 
Another significant effort in sentiment analysis on Twitter data is by Barbosa [16]. They use polarity predictions from three websites as noisy labels to train a model and use 1000 manually labelled tweets for tuning and another 1000 for testing. They propose the use of syntax features of tweets like punctuation, retweet, hashtags, link, and exclamation marks in addition with features like prior polarity of words and POS of words. \\
Some works leveraged the use of existing hashtags in the Twitter data for building the training data. (Davidov, Tsur, and Rappoport 2010) also use hashtags for creating training data, but they limit their experiments to sentiment/non-sentiment classification, rather than 3-way polarity classification, as [15] does. Our model integrates some of the preprocessing techniques this work used. Hassan Saif [9] introduced a novel approach of adding semantics as additional features into the training set for sentiment analysis. This approach works well for topic specific data. Hence, we thought of taking a different approach for a generic tweet dataset like ours.   

\section{Subjectivity}
Subjectivity refers to how someone's judgment is shaped by personal opinions and feelings instead of outside influences. An objective perspective is one that is not influenced by emotions, opinions, or personal feelings - it is a perspective based in fact, in things quantifiable and measurable. A subjective perspective is one open to greater interpretation based on personal feeling, emotion, aesthetics, etc. \\
Subjectivity classification is another topic in the domain of text classification which is garnering more and more interest in the field of sentiment analysis. Since a single sentence may contain multiple opinions and subjective and factual clauses, this problem is not as straightforward as it seems. Below are some examples of subjective and objective sentences. \\
\newline 
Objective sentence with no sentiment: So, the Earth revolves around the Sun. \\
Objective sentence with sentiment: The drug relieved my pain.\\
Subjective sentence with no sentiment: I believe he went home yesterday.\\
Subjective sentence with sentiment: I am so happy you got the scholarship.\\ 
\newline 
Classifying a sentence as subjective or objective provides certain conclusions. Purely objective sentences do not usually convey any sentiment, while most of the purely subjective sentences have a clear inclination towards either the positive or negative sentiment. Sentences which are not completely subjective or objective may or may not convey a sentiment. Libraries like TextBlob, and tools like Opinion Finder can be used to find the extent to which a sentence can be considered subjective.\\
Since tweets are usually person-specific, or subjective, we use this intuition to reduce the size of the training set by filtering the sentences with a subjectivity level below a certain threshold (fairly objective tweets).      

\section{Implementation}
In this section, we explain the various preprocessing techniques used for feature reduction, and also the additional step of filtering the training dataset using the subjectivity score of tweets. We further describe our approach of using different machine learning classifiers and feature extractors. We also propose an additional heuristic for sentiment classification which can be used as a tag-along with the learning heuristics. 
\subsection{Corpus}
Our training dataset\footnote{The URL is http://twittersentiment.appspot.com/. This page has a link to our training data and test data. It is also a public tool that other researchers can use to build their own data sets.} has 1.6 million tweets, and 5000 tweets in the test dataset. Since the test dataset provided comprised only 500 tweets, we have taken part of the training data (exactly 5000 tweets, distinct from the training dataset) as the test dataset. We remove emoticons from our training and test data. The table below shows some sample tweets.\\ 
\begin{center}
\begin{tabular}{|c|c|}
\hline
    \small Tweet & \small Sentiment \\
\hline
    \small @MrZeroo00 Yeah! tks man & \small Positive \\
    \small oh so bored...stuck at home & \small Negative\\
    \small pizza night and i feel too sick & \small Negative\\
\hline
\end{tabular}
\end{center}
\subsection{Subjectivity Filtering}
This is a new step we propose to achieve higher accuracy on a smaller training dataset. We use TextBlob to classify each tweet as subjective or objective. We then remove all tweets which have a subjectivity level/score (score lies between 0 and 1) below a specified threshold. The remaining tweets are used for training purposes. We observe that a considerable number of tweets are removed as the subjectivity threshold increases. We show the effect of doing this procedure on the overall accuracy in the evaluation section of the paper.
\subsection{Preprocessing}
The Twitter language model has many unique properties. We take advantage of the following properties to reduce the feature space. Most of the preprocessing steps are common to most of the previous works in the field. However, we have added some more steps to this stage of our model. 
\subsubsection{Basic steps}
We first strip off the emoticons from the data. Users often include twitter usernames in their tweets in order to direct their messages. We also strip off usernames (e.g. @Chinmay) and URLs present in tweets because they do not help us in sentiment classification. Apart from full stops, which are dealt in the next point, other punctuations and special symbols are also removed. Repeated whitespaces are replaced with a single space. We also perform stemming to reduce the size of the feature space.
\subsubsection{Full Stops}
In the previous works, full stops are just usually replaced by a space. However, we have observed that casual language in tweets is often seen in form of repeated punctuations. For example, ``this is so cool...wow". We take into consideration this format, and replace two or more occurrences of ``." and ``-" with a space. Also, full stops are also quite different in usage. Sometimes, there isn't any space in between sentences. For example, ``It’s raining.Feeling awesome". We replace a single occurrence of a full stop with a space to ensure correct feature incorporation.  
\subsubsection{Parsing Hashtags}
In the case of hashtags, most of the previous works just consider the case of hashtags followed by a single word; they just remove the hashtag and add the word to the feature vector. However, sometimes, there are multiple words after a hashtag, and more often than not, these words form an important, conclusive part of the Tweet. For example, \#ThisSucks, or \#BestMomentEver. These hashtags need to be dealt with in a proper fashion. We split the text after hashtags after before each capital letter, and add these as tokens to the feature vector. For hashtags followed by a single word, we just replace the pattern \#word with the word, as conventional models do. The intuition behind this step is that quite often, the sentiment of a tweet is expressed in form of a hashtag. For example, \#happy or \#disappointed are frequently used hashtags, and we don’t want to lose this information during sentiment classification. 
\subsubsection{Repeated letters}
Tweets contain very casual language as mentioned earlier. For example, if we search ``wow" with an arbitrary number of o's in the middle (e.g. wooow, woooow) on Twitter, there will most likely be a non-empty result set. We use preprocessing so that any letter occurring more than two times in a row is replaced with two occurrences. In the samples above, these words would be converted into the token ``woow". After all the above modifications, tweets are converted into lowercase to avoid confusion between features having same content, but are different in capitalization. 
\subsubsection{Stopwords, Acronyms and Negations}
We gather a list of 400 stopwords. These words, if present in the tweets, are not considered in the feature vector.  \\
We store an acronym dictionary which has over 5000, frequently-used acronyms and their abbreviations. We replace such acronyms in tweets with their abbreviation, since these can be of great use while sentiment classification.\\
All negative words like 'cannot', 'can't', 'won't', 'don't' are replaced by 'not', which effectively keeps the sentiment stable. It is observed that doing this makes the training faster, since the model has to deal with a smaller feature vector.  
\subsection{Baseline model}
The baseline model for our experiments is explained in the paper by Alec Go [1]. The model uses the Naive Bayes, SVM, and the Maximum Entropy classifiers for their experiment. Their feature vector is either composed of Unigrams, Bigrams, Unigrams + Bigrams, or Unigrams + POS tags. \\
This work achieved the following maximum accuracies:\\
a) 82.2 for the Unigram feature vector, using the SVM classifier,\\
b) 83.0 for the Unigram + Bigram feature vector, using the MaxEnt classifier, and 82.7 using the Naive Bayes classifier. \\
c) 81.9 for the Unigram + POS feature vector, using the SVM classifier.\\
These baseline accuracies were on a training dataset of 1.6 million tweets, and a test dataset of 500 tweets. We are using the same training dataset for our experiments. We later present the baseline accuracies on a training set of 200K tweets, and a test dataset of 5000 tweets; we compare our model's accuracy with these baseline accuracy values on the same test data of 5000 tweets.  
\subsection{Effective Word Score (EFWS) Heuristic}
We have described our baseline model above. So the feature vectors we collate results for, are Unigram, Unigram + Bigram, and Unigram + POS. We have already made two major changes before the training starts on our dataset as compared to our baseline model. Firstly, our training dataset will be filtered according to the subjectivity threshold. And secondly, our preprocessing is much more robust as compared to their work.\\
Now let us look at an additional heuristic we use to obtain labels for our test data. Along with dictionaries for stop words and acronyms, we also maintain a dictionary of a list of frequently used words and their polarity scores. This dictionary has around 2500 words and their polarity score ranging from -5 to 5. At runtime, we also use all synonyms of a word (from WordNet) present in a tweet and also the dictionary, and assign them the same score as the dictionary word. There is a reasonable assumption here, that the synonyms aren't very extremal in nature, that is, a word with a polarity score of 2 cannot have a synonym which has a polarity score of 5. Now, we calculate the Effective Word Scores of a tweet.\\ 
\newline
We define the Effective Word Score of score x as \\
~\\
\textit{EFWS(x) = N(+x) - N(-x)},\\
~\\
where N(x) is the number of words in the tweet with polarity score x.\\ 
\newline
For example, if  a tweet has one word with score 5, three words with score 4, two with score 2, three with with score -2, one with score -3, and finally two with score -4, then the effective word scores are:\\
\newline
EFWS(5) = N(5) - N(-5) = 1 - 0 = 1\\
EFWS(4) = N(4) - N(-4) = 3 - 2 = 1\\
EFWS(3) = N(3) - N(-3) = 0 - 1 = -1\\
EFWS(2) = N(2) - N(-2) = 2 - 3 = -1\\
EFWS(1) = N(1) - N(-1) = 2 - 0 = 2\\
\newline 
We now define the heuristic for obtaining the label of a Tweet. \\
\begin{algorithmic}
\IF {(EFWS(5) $\geq$ 1 or EFWS(4) $\geq$ 1) and (EFWS(2) $\geq$ 1)}
	\STATE Label = positive 
\ENDIF
\end{algorithmic}
~\\
Similarly, \\
\begin{algorithmic}
\IF {(EFWS(5) $\leq$ -1 or EFWS(4) $\leq$ -1) and (EFWS(2) $\leq$ -1)}
	\STATE Label = negative
\ENDIF
\end{algorithmic}
~\\
The basic intuition behind such a heuristic is that we found tweets having one strongly positive and one moderately positive word more than the number of strongly negative and the moderately negative words respectively, usually conveyed a positive sentiment. Similar was the case for negative sentiments. The tweets getting a label from this heuristic are not sent into the training phase. After considerable amount of experimenting, and analyzing the nature of our dataset, which is not domain specific, we have reached the conclusion that the heuristic mentioned above is optimal for obtaining labels. We found that the heuristic accuracy was around 85\% for a training dataset of 100K and a test dataset of 5K, where the total number of test tweets labelled by the heuristic were around 500. This means that around 425 out of the 500 tweets received a correct prediction of sentiment using this heuristic. \\
Thus, using this heuristic improves the overall accuracy, as well as saves time by reducing the number of tweets to be tested by the ML algorithms. 
\subsection{Training Model}
We use the following classifiers for our model. 
\subsubsection{Naive Bayes}
Naive Bayes is a simple model which works well on text categorization. We use a Naive Bayes model. Class c* is assigned to tweet d, where	c* = argmax P(c$|$d). \[P_{NB}(c|d) = P(c) * \sum_{i=1}^{m} P(f|c)^{n_i(d)}\] And $P_{NB}(c|d)$ is calculated using Bayes Rule. In this formula, f represents a feature and $n_i(d)$ represents the count of feature $f_i$ found in tweet d. There are a total of m features. Parameters P(c) and $P(f|c)$ are obtained through maximum likelihood estimates.
\subsubsection{Support Vector Machines}
Support vector machines are based on the Structural Risk Minimization principle from computational learning theory. SVM classification algorithms for binary classification is based on finding a separation between hyperplanes defined by classes of data. One remarkable property of SVMs is that their ability to learn can be independent of the dimensionality of the feature space. SVMs can generalize even in the presence of many features as in the case of text data classification. We use a non-linear Support Vector Machine with an RBF kernel.
\subsubsection{Maximum Entropy Model}
Maximum Entropy Model belongs to the family of discriminative classifiers also
known as the exponential or log-linear classifiers.. In the naive Bayes classifier, Bayes rule is used to estimate this best y indirectly from the likelihood $P(x|y)$ (and the prior $P(y)$) but a discriminative model takes this direct approach, computing $P(y|x)$ by discriminating among the different possible values of the class y rather than first computing a likelihood. \[ \hat{y} = \underset{y}{\operatorname{argmax}} P(y|x) \]
Logistic regression estimates $P(y|x)$ by combining the feature set  linearly (multiplying each feature by a weight and adding them up), and then applying a function to this combination.

\section{Evaluation}
In this section, we present the collated results of our experiments. To show that our model achieves higher accuracy than the baseline model and on a smaller training dataset, we first fix the test dataset. Our test dataset, as mentioned before, consists of 5000 tweets. We conducted our experiments on an Intel Core i5 machine (4 cores), with 8 GB RAM. The following are the accuracies of the baseline model on a training set of 200K tweets:

\begin{center}
\begin{tabular}{|p{2.2cm}|p{1.5cm}|p{1cm}|p{2.5cm}|}
\hline
     & \footnotesize Naive Bayes & \footnotesize SVM & \footnotesize Logistic Regression\\
\hline
    \footnotesize Unigram & 78.23\% & 74.10\% & 79.03\% \\
    \footnotesize Unigram + Bigram & 77.5\% & 71.3\% & 80.2\% \\
    \footnotesize Unigram + POS & 76.7\% & 71.8\% & 79.7\% \\ 
\hline
\end{tabular}
\end{center}

We filtered the training set with a subjectivity threshold of 0.5. By doing this, we saw that the number of tweets reduced to approximately 0.6 million tweets from an earlier total of 1.6 million. We then trained our model described in earlier sections on a 100K tweets randomly picked from this filtered training dataset, and observed the following accuracies:
\begin{center}
\begin{tabular}{|p{2.2cm}|p{1.5cm}|p{1cm}|p{2.5cm}|}
\hline
     & \footnotesize Naive Bayes & \footnotesize SVM & \footnotesize Logistic Regression\\
\hline
    \footnotesize Unigram & 79.2\% & 77.8\% & 80.5\% \\
    \footnotesize Unigram + Bigram & 77.9\% & 71.7\% & 81.7\% \\
    \footnotesize Unigram + POS & 77.5\% & 73.6\% & 79.9\% \\ 
\hline
\end{tabular}
\end{center}

Note that all the accuracies in the tables above have been recorded as the average of 3 iterations of our experiment. We achieve higher accuracy for all feature vectors, on all classifiers, and that too from a training dataset half the size of the baseline one. \\
\newline 
We now see the intricacies of the subjectivity threshold parameter. It is clear that more and more tweets get filtered as the subjectivity threshold parameter increases. This can be seen in the Figure 1 shown below. We have plotted the number of tweets that remain after filtering from two sources: TextBlob, Opinion Finder Tool\footnote{\*This tool can be found at: http://mpqa.cs.pitt.edu/opinionfinder/}. TextBlob has an inbuilt function that provides us the subjectivity level of a tweet. On the other hand, Opinion Finder only provides the information of which parts of the text are subjective, and which are objective. From that, we define the subjectivity level of that text as: \\
\newline
Subjectivity level = $\dfrac{\sum{\text{Length of subjective clauses}}}{\text{Total length of the text}}$ 

\begin{center}
\begin{tikzpicture}
\begin{axis}[
    xlabel={Subjectivity Threshold},
    ylabel={Tweets (in millions)},
    xmin=0, xmax=1,
    ymin=0, ymax=2000000,
    xtick={0,0.1,0.2,0.3,0.4,0.5,0.6,0.7,0.8,0.9,1},
    ytick={0,200000,400000,600000,800000,1000000,1200000,1400000,1600000,1800000},
    legend pos=north east,
]
\addplot[color=red]
    coordinates{
      (0, 1600000)
      (0.1, 939785)
      (0.2, 873054)
      (0.3, 804820)
      (0.4, 712485)
      (0.5, 571864)
      (0.6, 449286)
      (0.7, 304874)
      (0.8, 211217)
      (0.9, 135788)
    };
    
 \addplot[color=blue]
     coordinates{
       (0, 1600000)
       (0.1, 602313)
       (0.2, 499173)
       (0.3, 392223)
       (0.4, 262109)
       (0.5, 169477)
       (0.6, 154667)
       (0.7, 139613)
       (0.8, 126148)
       (0.9, 116842)
     };
     \legend{Textblob, Opinion Finder}    

\end{axis}    
\end{tikzpicture}
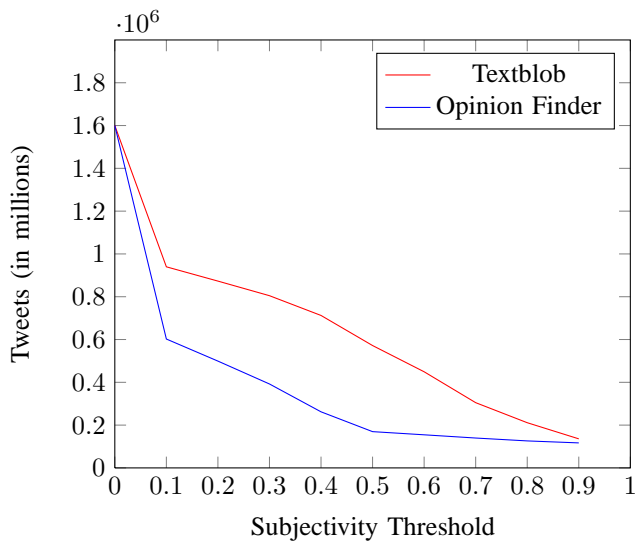
Figure 1: Number of tweets with subjectivity greater than the subjectivity threshold
\end{center}

\begin{center}
\begin{tikzpicture}
\begin{axis}[
    xlabel={Subjectivity Threshold},
    ylabel={Accuracy (from 0 to 1)},
    xmin=0, xmax=1,
    ymin=0.7, ymax=1,
    xtick={0,0.1,0.2,0.3,0.4,0.5,0.6,0.7,0.8,0.9,1},
    ytick={0,0.1,0.2,0.3,0.4,0.5,0.6,0.7,0.8,0.9,1},
    legend pos=north east,
]
\addplot[color=red]
  coordinates{
    (0.1, 0.753871866) 
    (0.2, 0.779442897)
    (0.3, 0.763421155) 
    (0.4, 0.783231198)
    (0.5,0.805132645)
    (0.6,0.807373259)
    (0.7,0.808587744)
    (0.8,0.817799443)
    (0.9,0.823872989)
};
\end{axis}
\end{tikzpicture}
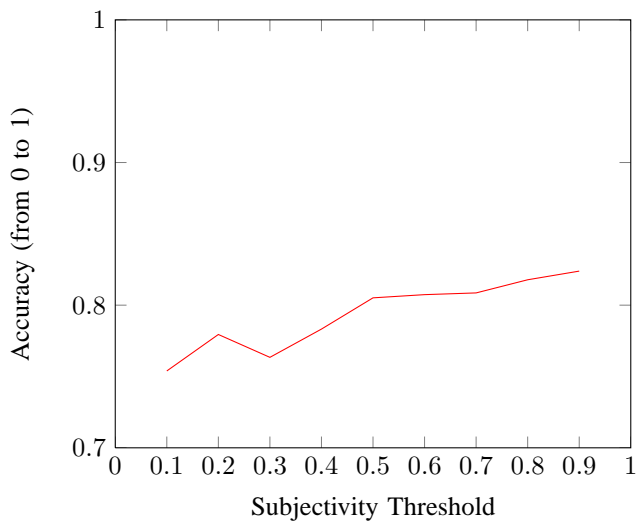
Figure 2: Variation of accuracy (*Training data of 100K, Test data of 5K) with subjectivity threshold. *TextBlob is used to filter the tweets to form the training dataset. 
\end{center}

We now focus on the issue of choosing the optimum threshold value. As the subjectivity threshold parameter increases, our model trains on tweets with a higher subjectivity level, and the overall accuracy increases. We observed the following accuracies on subjectivity level 0.8 (Unigrams as features):\\
Naive Bayes: 80.32\% \\
Non-linear SVM: 80.15 \% \\
Logistic Regression: 81.77\% \\
\newline 
We should consider the fact that a lot of useful tweets are also lost in the process of gradually increasing the parameter, and this could cause a problem in cases when the test data is very large, because the model will not train on a generic dataset. Researchers may use a higher subjectivity threshold for their experiments if they are confident that most of the important information would be retained. This is most likely to happen in case of topic-specific or domain-specific data.\\

\begin{center}
\begin{tikzpicture}
\begin{axis}[
    ybar,
    enlargelimits=0.15,
    legend style={anchor=north},
    legend pos= north east,
    ylabel={Training time (in minutes)},
    symbolic x coords={baseline,subjectivity=0.5,subjectivity=0.8},
    xtick=data,
    ]
\addplot coordinates {(baseline,17.4) (subjectivity=0.5,12.55) (subjectivity=0.8,10.68)};
\addplot coordinates {(baseline,16.23) (subjectivity=0.5,12.31) (subjectivity=0.8,10.34)};
\addplot coordinates {(baseline,31.9) (subjectivity=0.5,18.24) (subjectivity=0.8,16.3)};
\legend{Logistic Regression,Naive Bayes,SVM}
\end{axis}
\end{tikzpicture}

Figure 3: Comparison of training times for Unigrams
\end{center}

\begin{center}
\begin{tikzpicture}
\begin{axis}[
    ybar,
    enlargelimits=0.15,
    legend style={anchor=north},
    legend pos= north east,
    ylabel={Training time (in minutes)},
    symbolic x coords={baseline,subjectivity=0.5,subjectivity=0.8},
    xtick=data,
    ]
\addplot coordinates {(baseline,28.41) (subjectivity=0.5,14.09) (subjectivity=0.8,11.3)};
\addplot coordinates {(baseline,16.6) (subjectivity=0.5,13.51) (subjectivity=0.8,12.66)};
\addplot coordinates {(baseline,35.2) (subjectivity=0.5,20.6) (subjectivity=0.8,19.2)};
\legend{Logistic Regression,Naive Bayes,SVM}
\end{axis}
\end{tikzpicture}
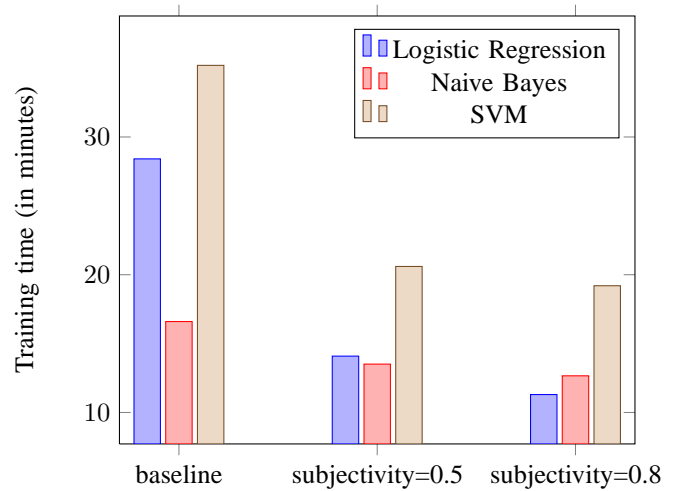
Figure 4: Comparison of training times for Unigrams + Bigrams
\end{center}

We use Logistic regression for classification and unigrams as the feature vector with K-fold cross validation for determining the accuracy. We choose an optimal threshold value of 0.5 for our experiment, considering the fact that the model should train on a more generic dataset. Figure 2 shows the variation of accuracy with the subjectivity threshold. The training size is fixed at 100K and the test dataset (5K tweets) is also same for all the experiments.\\ 
\newline 
We also measure the time taken to train our model, and compare it to the baseline model. Our observation was that our model took roughly half the amount of time in some cases and yet obtained a higher accuracy. Figures 3 and 4 show the difference in training time of the baseline model, our model on a 0.5 subjectivity-filtered dataset, and our model on a 0.8 subjectivity-filtered dataset on unigrams and unigrams + bigrams respectively. The times recorded are on a training dataset of 100K for our model and 200K for the baseline model, and a test dataset of 5K was fixed in all the recordings. The winning point, which can be seen from the plots, is that our model is considerably faster, and even has twofold speed in some cases. And alongside saving computation time, it achieves higher accuracy. This can be attributed to the fact that as the subjectivity threshold increases, only the tweets with highly polar words are retained in the training set and this makes the whole process faster.      

\section{Conclusion}
We show that a higher accuracy can be obtained in sentiment classification of Twitter messages training on a smaller dataset and with a much faster computation time, and hence the issue of constraint on computation power is resolved to a certain extent. This can be achieved using a subjectivity threshold to selectively filter the training data, incorporating a more complex preprocessing stage, and using an additional heuristic for sentiment classification, along with the conventional machine learning techniques. As Twitter data is abundant, our subjectivity filtering process can achieve a better generalised model for sentiment classification.





%

\end{document}